\documentclass[amsmath,amssymb,preprint,showpacs]{revtex4}
\usepackage{graphicx}
\begin{document}

\title{Negative refraction in nonlinear wave systems}

\author
{Zhoujian Cao$^{1}$, Hong Zhang$^{2}$, and Gang
Hu$^{1}$\footnote{Author for correspondence: ganghu@bnu.edu.cn}}
\affiliation{$^{1}$Department of physics, Beijing Normal
University, Beijing,100875, China} \affiliation{$^{2}$Zhejiang
Institute of Modern Physics and
Department of Physics, Zhejiang\\
University, Hangzhou 310027, China}
\date{\today}

\begin{abstract}
People have been familiar with the phenomenon of wave refraction
for several centuries. Recently, a novel type of refraction, i.e.,
negative refraction, where both incident and refractory lines
locate on the same side of the normal line, has been predicted and
realized in the context of linear optics in the presence of both
right- and left-handed materials. In this work, we reveal, by
theoretical prediction and numerical verification, negative
refraction in nonlinear oscillatory systems. We demonstrate that
unlike what happens in linear optics, negative refraction of
nonlinear waves does not depend on the presence of the special
left-handed material, but depends on suitable physical condition.
Namely, this phenomenon can be observed in wide range of
oscillatory media under the Hopf bifurcation condition. The
complex Ginzburg-Landau equation and a chemical reaction-diffusion
model are used to demonstrate the feasibility of this nonlinear
negative refraction behavior in practice.
\end{abstract}


\maketitle

It is well known that light can refract at interface of two media.
For several centuries only positive refraction has been found in
nature, where incident line and refractory line locate on either
side of the normal line of the interface. Very recently, a new
type of optical refraction--negative refraction, where both
incident and refractory lines locate on the same side of the
normal line of the interface, has been theoretically predicted
\cite{veselago} and experimentally observed \cite{smith} when
refraction occurs between a right-handed material and a
left-handed material where optical waves propagate inward to the
wave source. In addition, some new observations have been reported
in nonlinear wave propagation, and one particularly interesting
phenomenon is antispiral waves
\cite{vanag1,vanag2,yang,gong,bar,kim}. In contrast to spirals
(which are familiar to people) where waves propagate outwards from
the tips, antispiral  waves propagate inward to the spiral tip,
the wave source. Because the existence of inwardly propagating
waves is key to generate negative refraction in linear optics, we
expect that the novel phenomenon of negative refraction may also
be observed in other nonlinear wave systems where antispirals and
inwardly propagating waves can be found.

In this letter, we go further from the above phenomena of negative
refraction in linear optics and antispirals in nonlinear waves
propagation by revealing a new type of negative refraction in
nonlinear oscillatory systems. We demonstrate that unlike what
happens in linear optics, negative refraction of nonlinear waves
does not require special left-handed materials, rather, it can be
found in wide range of media under suitable nonlinear condition of
Hopf bifurcation. To demonstrate this new phenomenon, we use the
complex Ginzburg-Landau equation (CGLE) as our example. With this
model, we are able to freely generate, by periodic forcing, both
normal waves (NWs) and antiwaves (AWs), of which NWs propagate
outward from the wave source while AWs propagate towards the wave
source. An interesting feature of negative refraction is observed
when waves on either side of the interface have different
propagation characteristics, one AW and the other NW. Moreover, we
use a typical chemical reaction diffusion model, the Brusselator,
to verify the prediction of negative refraction of CGLE. This
suggests possible experimental realization of negative refraction
in nonlinear chemical reaction diffusion systems.

We consider 2D CGLE with local periodic
forcing \cite{zhang}
\begin{eqnarray}
\frac{\partial A}{\partial
t}=A-(1+i\alpha)|A|^{2}A+(1+i\beta)\nabla^{2}A+\delta(x)Fe^{-i\omega_{in}
t}\label{cgle}
\end{eqnarray}
where no-flux (Neumann) condition is applied to all boundaries.
The last term on the right side of the equation, an additive
periodic forcing with constant amplitude $F$, serves as a wave
source applied to the $x=0$ stripe of the medium. $F$ needs to be
sufficiently large so that periodic waves can be stimulated, which
will then propagate as planar waves into the unforced region of
$x>0$. We use $F=1$ throughout this paper. The forcing frequency
$\omega_{in}$ has its sign, which denotes the rotating direction
of the forcing in the complex plane, ``+" clockwise and ``-"
anticlockwise. In the absence of the source term ($F=0$), Eq.(1)
describes the onset of homogeneous oscillation through Hopf
bifurcation, and this equation is used extensively as a typical
model for describing rich behaviors of spatiotemporal chaos,
propagating waves and pattern formations
\cite{7,89,1011,12,13,14}.

In Fig.1, we fix $\beta=-1.4$ and demonstrate distinctive
dynamical behaviors of the systems for different $\alpha$ and
$\omega_{in}$. Some parameter combinations generate NW (Fig.
1(a)), while others generate AW (Fig. 1(b)). The parameter regions
in Fig. 1(c) indicated by NW and AW show NW and AW outputs,
respectively, with $\omega=\omega_{in}$ ($\omega$ denotes the
output frequency). In the shadowed region, the periodic forcing
cannot control the system, and thus the system motion under
periodic forcing is either periodic with $\omega\neq\omega_{in}$,
or chaotic with broad band frequency spectrum. In this study we
are not interested in the shadowed region, and will focus on the
NW and AW regions only where $\omega=\omega_{in}$. Interestingly,
all the NW and AW regions locate in the vicinity of the diagnol
line $\omega=\omega_{in}=\alpha$. Since $\alpha$ is the bulk
oscillation frequency ($\omega_{o}=\alpha$) of the local dynamics,
all the NW and AW regions of Fig.1(c) satisfy resonance condition
($\omega_{in}\approx\omega_{o}=\alpha$). Moreover, in Fig. 1(c) it
is clear that AWs appear only under the condition
\begin{figure}
\includegraphics[ height = 4.0in, width = 0.6\linewidth]{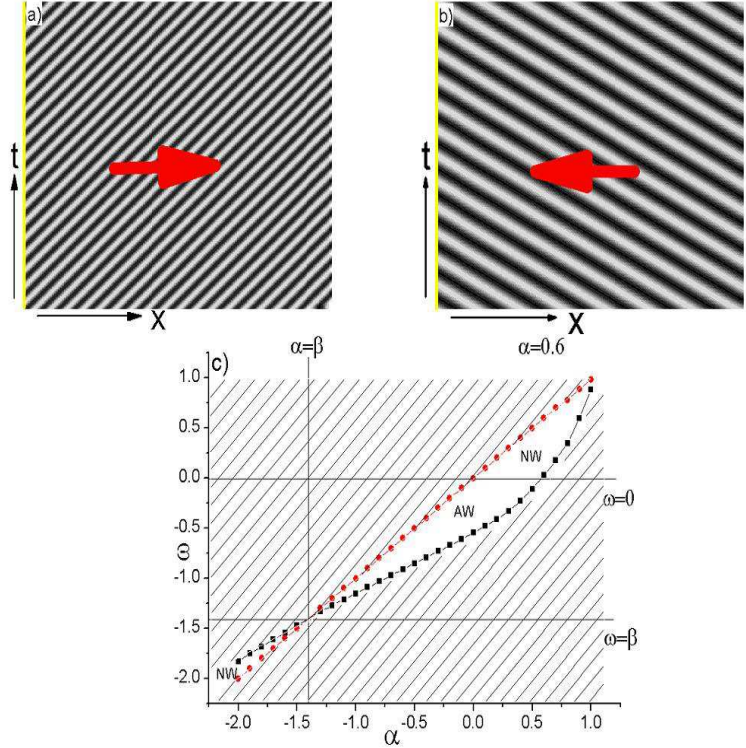}
\caption{In this and all the following figures, we apply no-flux
(Neumann) boundary condition. $\beta=-1.4$, $F=1.0$ are used in
Figs.1 and 3 and $256\times 256$ rectangular geometry is used for
numerical simulations in Fig.1. (a \& b) Numerical results of
spatiotemporal patterns at $y=128$, $x\in [0,256]$ with different
periodic forces applied on the left boundary of the rectangular
system. Arrows denote the wave propagation directions. Space
descritization of $256\times256$ pixes and time step $\Delta
t=0.05$ are used for numerical simulations. (a): Normal forward
propagating waves (NWs). $\alpha=0.2$, $\omega=\omega_{in}=-0.15$.
(b): Planar antiwaves whose phase propagates toward the wave
source. $\alpha=0.2$, $\omega=\omega_{in}=0.15$. (c):
Distributions of different characteristic regions in
$\omega_{in}-\alpha$ plane. Shadowed region: no stable waves with
$\omega=\omega_{in}$ can be realized in Eq.(1) whatever the input
frequency $\omega_{in}$. NW white region: normal waves of
$\omega=\omega_{in}$ can be observed; AW white region: antiwaves
of $\omega=\omega_{in}$ can be observed.}
\end{figure}
\begin{eqnarray}
\omega\omega_{0}>0&\text{and}&|\omega|<|\omega_{0}|=|\alpha|
\end{eqnarray}
NWs can be observed under the conditions of $\omega\omega_{0}<0$
or $|\omega|>|\omega_{0}|$ for $\omega\omega_{0}>0$. The
boundaries between the parameter domains of NW and AW can be
determined by the sign change of the phase velocity \cite{bar} $
v_{phase}=\frac{\omega}{k}$. The transitions between NW and AW
thus occur at two boundaries: $\omega=0$ and $k=0$. The former
boundary justifies the transition between AW and NW with
stationary Turing patterns $\omega=\omega_{in}=0$ at the
transition, and the latter defines the transition between AW and
NW with homogeneous oscillation ($k=0$) at the transition, which
shrinks to a single point of $\alpha=\beta=\omega_{in}$ in
Fig.1(c). In summary, we find that, with local forcing method, we
can generate NWs and AWs freely. For instance, given an
oscillatory medium with fixed parameters, we can produce both NW
and AW by changing the forcing frequency $\omega_{in}$ only. This
is different from antispirals in autonomous systems which produce
either NW or AW (not both) with a single spiral frequency selected
at the given set of parameters.

\begin{figure}
\includegraphics[ height = 2.0in, width = 0.5\linewidth]{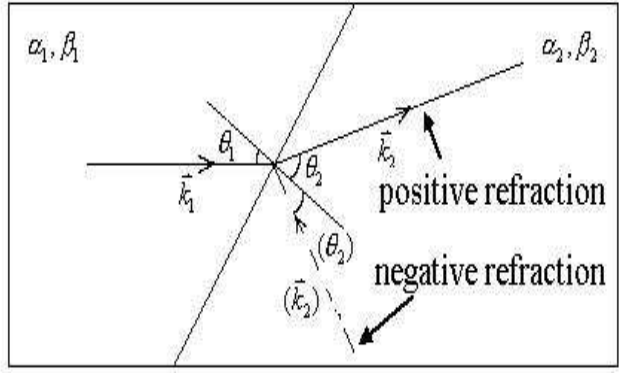}
\caption{Schematic figure of nonlinear refraction. The left and
right regions are CGLE systems with parameters $\alpha_{1}$,
$\beta_{1}$ and $\alpha_{2}$, $\beta_{2}$, respectively. The
middle line from left below to right up is the interface. The line
marked with $\vec{k}_{1}, \vec{k}_{2},(\vec{k}_{2})$ are incident
line, positive refractory line, and negative refractory line,
respectively.}
\end{figure}

Now we come to the problem of nonlinear wave refraction. We
consider planar waves passing an interface of two media with
different parameters $\alpha$, $\beta$ (Fig. 2). Analogous to
optics, the incident line and refractory line are perpendicular to
the wave fronts in the two media. The incident angle
($\theta_{1}$) is defined as the angle between the normal line and
the incident line, which is positive definite. The refractory
angle ($\theta_{2}$) is defined as the angle between the normal
line and the refractory line, which is positive if the refractory
and incident lines locate on either side of the normal line, and
is negative otherwise.

In linear optics, the light refraction relation can be precisely
computed, based on the continuity conditions at the interface. For
nonlinear systems, there is no general theory to describe the law
of nonlinear wave refraction although the continuity conditions
are still valid at the interface. The difficulty is that no
analytical solutions can be obtained around the interface of two
nonlinear media. We approximate the solutions of Eq.(1) as planar
wave solutions \cite{aranson} \setcounter{equation}{0}
\renewcommand{\theequation}{3\alph{equation}}
\begin{eqnarray}
A_{1,2}&=&\sqrt{1-k_{1,2}^{2}}e^{i(-\omega_{1,2}t+\vec{k}_{1,2}\cdot\vec{r})}\\
\omega_{1,2}&=&\alpha_{1,2}+(\beta_{1,2}-\alpha_{1,2})k^{2}_{1,2}
\end{eqnarray}
in the left and right regions of Fig.2, respectively. The actual
solutions around the interface certainly deviate from these planar
wave solutions. For instance, the amplitude of $|A_{1}|$ and
$|A_{2}|$ must change considerably in the interface region for
satisfying the continuity conditions. Our assumption is that the
interface causes only very weak influence to the phase
distributions of the planar waves (this assumption is fully
confirmed by our numerical simulations). This assumption allows us
to predict the nonlinear wave refraction law by applying the phase
continuity hypothesis of the planar wave solutions on the
interface: \setcounter{equation}{3}
\renewcommand{\theequation}{\arabic{equation}}
\begin{eqnarray}
-\omega_{1}t+\vec{k}_{1}\cdot\vec{r}=-\omega_{2}t+\vec{k}_{2}\cdot\vec{r}
\end{eqnarray}
Here, we consider the case of $\omega_{1}=\omega_{2}$ only,
otherwise the refractory pattern cannot be stationary \cite{ott}.
Therefore, we have
$\vec{k}_{1}\cdot\vec{r}=\vec{k}_{2}\cdot\vec{r}$ which leads to $
k_{1}\sin{\theta_{1}}=k_{2}\sin{\theta_{2}}$.

There are several interesting conclusions we can draw from Eq.(4).
First, with the sign of $\omega$ fixed (e.g., positive $\omega$)
$k_{1}$ and $k_{2}$ can be positive (NW) or negative (AW). It
tells us that if $k_{1}$ and $k_{2}$ have the same sign positive
refraction will be observed. If $k_{1}$ and $k_{2}$ have different
signs, i.e., from NW at left to AW at right or vise versa,
nonlinear negative refraction, an analog of propagation of light
from right- to left-handed material (or vice versa), emerges.
Second, with the help of dispersion relation, we can define from
Eq.(3b) an effective refractory index similar to that in optics as
\begin{eqnarray}
n(\omega)=\pm\frac{1}{|\omega|}\sqrt\frac{\omega-\alpha}{\beta-\alpha}
\end{eqnarray}
where ``+'' for NW ($\omega\omega_{0}<0$ or
$|\omega|>|\omega_{0}|=|\alpha|$ and $\omega\omega_{0}>0$) and
``-'' for AW ($\omega\omega_{0}>0$ and $|\omega|<|\omega_{0}|$).
The generalized refraction law for nonlinear waves in CGLE can be
expressed as
\begin{eqnarray}
\frac{\sin\theta_{1}}{\sin\theta_{2}}=\frac{n_{2}}{n_{1}}
\end{eqnarray}

In the following, we conduct numerical simulations of (1) to test
all the above analytical results. In our simulations, we divide a
rectangular region into two parts like Fig. 2.  In Fig. 3(a), a
pattern of positive refraction of nonlinear waves is illustrated
clearly with direct numerical simulation of Eq.(1). In Fig. 3(b),
we plot the relation of incident and refractory angles in
$\theta_{1}-\theta_{2}$ plane. The circles are numerical data and
the solid line is theoretical prediction, which perfectly coincide
with each other, fully confirming the assumption of Eq.(4) and the
prediction of (6).

Now we demonstrate negative refraction problem by simulation. In
Fig.3(c) and 3(d), we take parameter combinations at which the
incident waves are NWs while the refractory waves are AWs. Fig.
3(c) shows the pattern of negative refraction analogous to the
novel negative refraction in linear optics \cite{smith}. Unlike
the optical case, here the negative refraction is observed with
nonlinear waves. In Fig. 3(d), we perform the same computation as
Fig. 3(b) by considering negative refraction, and numerical
results are in perfect agreement with the theoretical predictions
for negative refraction.

\begin{figure}
\includegraphics[ height = 4.2in, width = 0.8\linewidth]{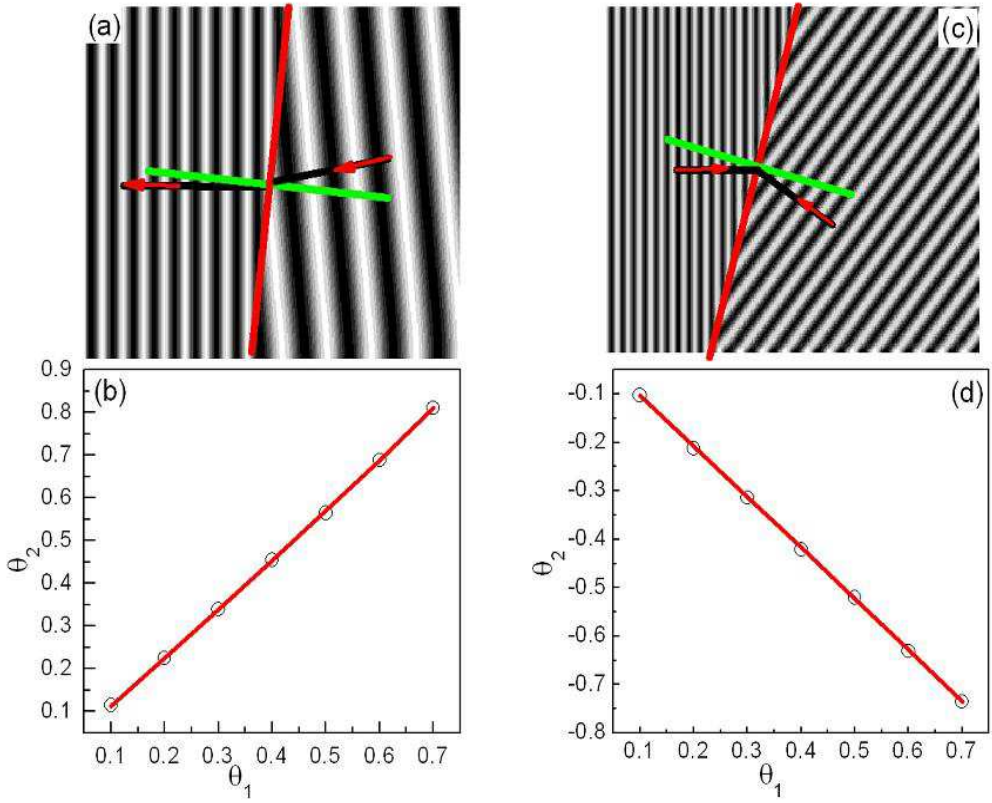}
\caption{(a \& b) Positive refraction. (a): $\omega_{in}=0.2$,
$\theta_{1}=0.1$. Pattern of positive refraction with
$\alpha_{1}=0.5$, $\beta_{1}=0.0$ (AW medium) and
$\alpha_{2}=0.5$, $\beta_{2}=-1.8$ (AW medium). The red line is
the interface of two media. The blue line is the normal line
perpendicular to the interface. The other two lines marked with
arrows are the incident and the refractory lines, respectively.
The arrows denote the directions of phase velocities. (b): The
relation between the incident angle ($\theta_{1}$) and the
refractory angle ($\theta_{2}$). $\alpha_{1}=0.5$,
$\beta_{1}=-1.0$, $\alpha_{2}=0.5$, $\beta_{2}=-1.4$, and
$\omega_{in}=-0.02$. Circles represent numerical results and the
solid line shows theoretical predictions of Eq.(6). Both results
coincide with each other perfectly. (c)\&(d) $\omega_{in}=0.02$.
Negative refraction with $\alpha_{1}=-0.5$, $\beta_{1}=1.4$ (NW
medium) and $\alpha_{2}=0.5$, $\beta_{2}=-1.4$ (AW medium). (c):
Pattern of negative refraction with $\theta_{1}=0.5$. (d): The
same as (b) with negative refraction considered.}
\end{figure}

Since CGLE appears universally in nonlinear extended systems
around Hopf bifurcation from homogenous stationary states to
homogeneous oscillations, the negative refraction phenomenon found
in CGLE is expected to be observed experimentally in wide range of
practical nonlinear and nonequilibrium systems. Chemical reaction
diffusion (RD) systems are the most probable candidates of this
kind. Here we take a well known chemical reaction model, the 2D
Brusselator, as our example \cite{Nicola,Nicolis}
\begin{eqnarray}
\frac{\partial u}{\partial t}&=&a-(b+1)u+u^{2}v+\nabla^{2}u+\delta(x)F\cos(\omega_{R} t)\nonumber\\
\frac{\partial v}{\partial t}&=&bu-u^2v+D\nabla^{2}u
\end{eqnarray}
In these equations, $a$ and $b$ are two constant chemical
components, and $u$ and $v$ chemical variable components. All $a$,
$b$, $u$ and $v$ are positive definite (physical constraint).
Without periodic forcing, this system undergoes Hopf bifurcation
to homogeneous oscillation at the parameters $b=1+a^2$. At the
Hopf instability, the frequency of the oscillation reads
$\omega_{H}=a$. We simulate Eq.(7) with a parameter set slightly
beyond Hopf bifurcation, and find positive refraction phenomenon
shown in Fig. 4(a). In Fig.4(b), we use $\omega_{R}=1.05$ and
parameter sets of $a=1.0$, $D=0.5$ and $a=1.2$, $D=2.25$ for the
two chemical media on either side of the interface, respectively.
The chemical parameter set can be transformed to CGLE parameter
set by $\alpha=a\frac{D-1}{D+1}$,
$\beta=\frac{7a^2-4-4a^4}{3a(2+a^2)}$ and
$\omega_{in}=\frac{\omega_{R}-a}{\sqrt{\epsilon}},
\epsilon=b-(1+a^2)$ \cite{Nicola}. From this correspondence, we
justify NW and AW for the left and right media of Fig.4(b),
respectively, and expect to observe negative refraction. Indeed,
we do reveal this interesting phenomenon in the simulation of this
nonlinear RD system in Fig.4(b).

\begin{figure}
\includegraphics[ height = 3.0in, width = 1.0\linewidth]{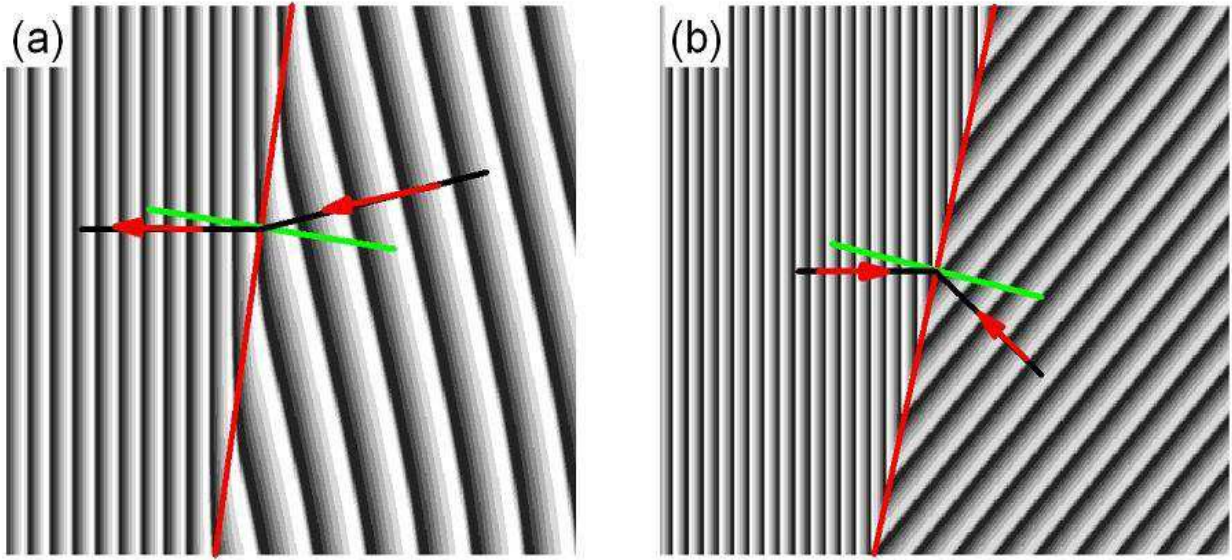}
\caption{Numerical simulation of Eq. (7), the periodically forced
Brusselator. Time step $\Delta t=0.01$. $400\times300$ physical
space is descritized with $400\times300$ pixes and the forcing is
applied on the sites of the left line of the rectangular space.
$b=3.2$. (a) Positive refraction observed between two AW media.
$a=1.2$, $D=1.7$ for the left medium and $a=1.2$, $D=2.5$ for the
right one. $\omega_{R}=1.07$, $\theta_{1}=0.2$. (b) Negative
refraction observed between NW (left) and AW (right) media.
$a=1.0$, $D=0.5$ for the left medium and $a=1.2$, $D=2.25$ for the
right one. $\omega_{R}=1.05$, $\theta_{1}=0.3$.}
\end{figure}

In conclusion, we have generated antiwaves in CGLE and reaction
diffusion systems by local periodic forcing. At fixed parameters,
we can freely produce normal or anti waves by adjusting the input
frequency $\omega_{in}$. A frequency dependent refraction index
$n(\omega)$ is defined for CGLE, which takes positive and negative
values for NW and AW, respectively. When waves propagate from
positive to negative $n(\omega)$ or vice versa, novel negative
refraction emerges. The negative refraction phenomenon predicted
in linear optics depends on the existence of left-handed materials
which have not yet found so far in nature. In our case, nonlinear
negative refraction can be observed in wide range of natural
oscillatory media under suitable physical condition, i.e., Hopf
bifurcation and resonance conditions. Since the complex
Ginzburg-Landau equation and nonlinear waves are common in nature,
we predict that the negative refraction phenomenon explored in
this letter can be widely observed, and that it can be realized in
experiments, such as in chemical reaction experiments.

\end{document}